\documentclass[pra,aps,twocolumn,showpacs]{revtex4}
\usepackage{graphicx}

\begin{document}

\title{Quasi-condensate Growth on an Atom Chip}
\author{N.P. Proukakis$^{1}$\footnote{Present Address: School of Mathematics and Statistics, University of
Newcastle, Merz Court, Newcastle NE1 7RU, United Kingdom},
J. Schmiedmayer$^{2}$ and H.T.C. Stoof$^{1}$}
\affiliation{$^1$Institute for Theoretical Physics, Utrecht University, Leuvenlaan 4, 3584 CE Utrecht,
The Netherlands}
\affiliation{$^2$Physikalisches Institut, Universit\"{a}t Heidelberg, 69120 Heidelberg, Germany}

\begin{abstract}

We discuss and model an experiment to study quasi-condensate growth
on an atom chip. In particular, we consider the addition of a deep dimple
to the weak harmonic trap confining an ultracold
one-dimensional atomic Bose gas,
below or close to the characteristic temperature for quasi-condensate formation.
The subsequent dynamics depends critically on both the initial conditions,
and the form of the perturbing potential.
In general, the dynamics features
a combination of shock-wave propagation in the quasi-condensate,
and quasi-condensate growth
from the surrounding thermal cloud.

\end{abstract}

\pacs{03.75.Kk, 05.30.Jp, 03.75.-b}

\maketitle


\section{Introduction}

The recent achievement of effectively one-dimensional ultracold atomic samples
on microfabricated surfaces, known as `atom chips' \cite{Atom_Chip_Review}
opens up the way for a number of
applications, such as precision interferometric
measurements \cite{Chip_Interf_1,Chip_Interf_2,Chip_Interf_3,Chip_Interf_4}
and quantum computation \cite{Chip_QC_1,Chip_QC_2,Chip_QC_3}.
One-dimensional (1D) geometries, in which the transverse confinement exceeds all
other relevant energy scales in the system, offer improved atomic guidance
and device miniaturization \cite{Real_1D_1,Real_1D_2}. This comes
at the expense of increased phase fluctuations, which tend to
destroy the coherence of the ensemble.
Various equilibrium studies have already been
performed both experimentally \cite{Low_D_Exp_1,Low_D_Exp_2,Low_D_Exp_3,Low_D_Exp_4}
and theoretically
\cite{Low_D_Theory_1,Low_D_Theory_2a,Low_D_Theory_2b,Low_D_Theory_2c,Low_D_Theory_2d,Low_D_Theory_3a,Low_D_Theory_3b,Low_D_Theory_3c,Low_D_Theory_3d,Low_D_Theory_3e}
to understand in particular this latter
effect. However,
a deeper understanding of the intrinsic dynamics of such systems is still lacking,
including the issue of quasi-condensate formation, and the role of phase fluctuations
on its growth dynamics.
Although condensate formation has been studied in 3D geometries
\cite{Ketterle_Growth,Dimple_Growth_1,3D_Growth,3D_Growth_Theory_1,3D_Growth_Theory_2,3D_Growth_Theory_3},
such studies are only now possible in 1D.
In this paper, we discuss a realistic experiment which captures the
dynamics of quasi-condensate formation and relaxation into a perturbed
potential, by adding
a dimple microtrap on an atom chip containing a gas of 1D ultracold Bose atoms.

The issue of quasi-condensate formation may be related to a recent
3D condensate growth experiment \cite{3D_Growth}, which to date remains largely not
understood.
In particular, this experiment revealed, under conditions of slow cooling,
  an unexpected slow linear initial
condensate growth.
As these features were observed close to the transition
point, the authors suggested this might be due
to the enhanced phase fluctuations in this region, leading to
the formation of a quasi-condensate that preceded the usual condensate growth.
The dimensionality of our envisaged 1D experiment ensures the system
remains
in the regime of large phase fluctuations throughout its entire evolution.
Moreover,
under conditions of slow cooling from above the transition point,
our results point towards a slow initial linear-like
growth which is consistent with bosonic
enhancement. However, a
detailed comparison to the above experiment would
require a full 3D calculation, and therefore lies beyond the scope of
this paper.

\begin{figure}[b]
\includegraphics[width=8.0cm]{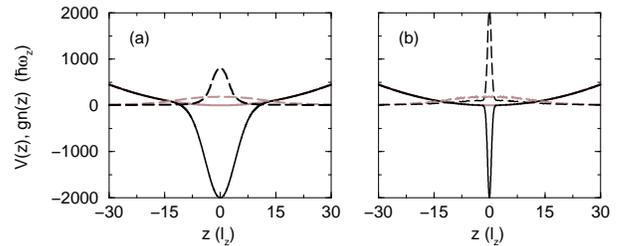}
\caption{
Initial (brown lines) and final (black) longitudinal harmonic potentials (solid)
and total densities (dashed) for a system of 2500 $^{87}$Rb atoms at a temperature $T=50 nK$
for (a) a wide dimple  $w=4l_{z}$, and (b) a tight dimple  $w=0.5l_{z}$, with
$z_{D}=0$, $V_{0}=-2000 \hbar \omega_{z}$. Here,
 $l_{z}=\sqrt{\hbar/m \omega_{z}}$ is the longitudinal harmonic oscillator length.
}
\end{figure}

The experimental scenario considered here is the following: An atom chip
is first loaded with a certain number of atoms, at densities low enough to be
in the 1D regime, and cooled to a prescribed low temperature.
The approximately
harmonic longitudinal confinement of the atoms is then perturbed by the addition of a deep dimple
microtrap, of variable width, as shown in Fig. 1.
Such a technique of local phase-space compression \cite{Compression_1,Compression_2}
has been used to force a system of $^{23}$Na atoms into the quantum-degenerate
regime in a reversible manner in 3D \cite{Dimple_Growth_1}.
A related approach has been used to create quantum-degenerate $^{133}$Cs in 3D
\cite{Dimple_3D_Cs_Exp,Dimple_3D_Cs_Th}, and in 2D \cite{2D_Dimple}.
In 1D, such a dimple can be created by
optical and magnetic traps, although
the optimum technique on an atom chip appears to be the application of electric fields
to a separate electrode, as discussed in Appendix A.
We study the resulting growth dynamics focusing on the regime of a very deep dimple, whose potential
greatly exceeds
all other energies in the system \cite{Prelim_Exp}, as this leads to the most interesting
non-equilibrium regime.
If there is {\rm already} a quasi-condensate present in the
original trap,
the addition of the dimple is found to lead to
shock-wave formation and subsequent relaxation
of the perturbed quasi-condensate
in the dimple.
In the opposite case of an initially {\rm incoherent} ultracold atomic sample,
we observe {\rm direct} quasi-condensate growth from the surrounding
thermal cloud.
The nature of the dynamics additionally depends on the width
of the dimple compared to the size of the atomic cloud,
and on the dimple location.
The distinct regimes of quasi-condensate dynamics
can be probed by controlling the initial temperature, atom number and perturbing potential.

This paper is structured as follows: Sec. II discusses the methodology used to model the envisaged
experiment. The dynamics following the addition of the dimple
in the limit of low temperature
is then discussed in Sec. III, paying
particular attention to shock-wave formation (Sec. III A),
its effect on the dynamics of the density at the trap centre (Sec. III B),
and the effect of the width of the dimple compared to the size of the atomic cloud (Sec. III C).
The effect of temperature is discussed in Sec. IV for both wide and tight dimples, with the latter
case highlighting the interplay between spontaneous and stimulated growth in the dimple.
Sec. V summarizes the examined regimes by means of
a suitable graph, and Sec. VI features a brief summary.
The discussion of experimental techniques which can be used to create the dimple trap
on an atom chip has been deferred to Appendix A.


\section{Methodology}

To model the proposed experiment, we first prepare
the desired
initial state on the atom chip stochastically.
This is achieved by means of the Langevin
equation
\begin{eqnarray}\nonumber i \hbar \frac{ \partial \Phi(z,t) }{ \partial t}
&= & \Bigg[ - \frac {\hbar^{2} \nabla^{2} }{2m} + V_{\rm ext}(z) - \mu
- iR(z,t)
\\& &
+ g |\Phi(z,t)|^{2} \Bigg] \Phi(z,t) + \eta(z,t)\;,
\label{lang}
\end{eqnarray}
describing the dynamics of the order
parameter $\Phi(z,t)$~\cite{Stoof_Noisy_1,Stoof_Noisy_2,Stoof_Noisy_3}.
Here $V_{\rm ext}(z)=m \omega_{z}^{2} z^{2}/2$ denotes the longitudinal harmonic confinement,
and $g=2 \hbar a \omega_{\perp}$
 is the one-dimensional coupling constant obtained by averaging over transverse
gaussian wavefunctions, where $a$ is the three-dimensional
scattering length and $\omega_{\perp}$ the trap frequency in the transverse
directions \cite{Olshanii}.
The chemical potential, $\mu$,
determines, for
a given initial trap potential and temperature, the total atom number in the system.
The atom chip trap is pumped from a
thermal reservoir at a rate \cite{Stoof_Noisy_1,Stoof_Noisy_2,Stoof_Noisy_3}
\begin{eqnarray}
iR(z,t)&=&-{\beta\over4}\hbar\Sigma^{\rm K}(z)\nonumber
\\&&
\hspace{-2cm}
\times
\left(
-{\hbar^2\nabla^2\over2m}+V^{\rm ext}(z)-\mu+g|\Phi(z,t)|^2
\right)\;.
\end{eqnarray}
In accordance with the
fluctuation-dissipation theorem, there is an associated gaussian noise
contribution $\eta(z,t)$ obeying
$\langle
\eta^*(z,t)\eta(z^\prime,t^{\prime})
\rangle
=(i\hbar^2/2)\Sigma^{\rm K}(z)\delta(z-z^\prime)\delta(t-t^{\prime})$,
where the brackets denote averaging over the realizations
of the noise.
The dependence of these quantities on the Keldysh
self-energy $\hbar\Sigma^{K}(z)$ ensures that the
trapped gas relaxes to the correct thermal equilibrium,
as additionally verified by direct comparison to the modified Popov
theory discussed elsewhere
\cite{Low_D_Theory_2a,Low_D_Theory_2b,Low_D_Theory_2c}.
For simplicity, we choose in this work both the longitudinal and the
transverse atom chip confinement to be harmonic, with respective trap frequencies
$\omega_{z} = 2 \pi \times 5$ Hz and $\omega_{\perp} = 2 \pi \times 5000$ Hz, and
present explicitly results for $^{87}{\rm Rb}$, with $a=5.32$nm.
Unless otherwise specified, results will be given in
dimensionless harmonic oscillator units of the original trap,
such that lengths are scaled to the longitudinal harmonic oscillator length
$l_{z} = \sqrt{\hbar / m\omega_{z}}$ and energies to $\hbar \omega_{z}$.

Once the system has fully relaxed to the desired equilibrium, we suddenly turn on a
gaussian dimple potential and simultaneously remove the coupling to the heat bath.
Switching off driving and dissipation terms, i.e., taking $R(z,t)=\eta(z,t)=0$,
ensures that the total atom number on the atom chip remains fixed,
as in the envisaged experiment,
and thus only intrinsic dynamics of the system are taken into account.
The dimple potential added is given by
$V_{D} = V_0 e^{-(z-z_D)^{2}/2w^{2}}$
where $V_{0}$ is the dimple depth, $w$ its width, and $z_{D}$ its location.
A sudden introduction of the dimple potential is chosen,
since this generates
a highly non-equilibrium situation, which gives rise to the most interesting dynamics.
For the chosen parameters, this corresponds to a trap turn-on timescale of $6 \mu$s, which is
an experimentally realistic timescale.
The subsequent analysis is based on averaging over many
different initial realizations, which ensures that both density and phase fluctuations are
accurately included.
This is equivalent to averaging over a large number of independent experimental realizations with
a variable initial phase, as typically done when investigating growth
dynamics  \cite{Ketterle_Growth,Dimple_Growth_1,3D_Growth}.
In our discussion, we assume the dimple is
turned on at $t=0$.

\section{Low Temperature Limit}

The sudden addition of a deep gaussian dimple on a harmonically confined pure condensate
leads to
a large atomic flux towards the center of the dimple,
and thus to the development of a large local density gradient at symmetric
points about the dimple centre,
which, in turn, leads to the
formation of two counter-propagating shock wavefronts.
The dynamics of the initial shock wave formation stage for a pure coherent condensate
has been discussed under various related conditions in Refs.
\cite{Shock_Waves_1,Shock_Waves_2,Shock_Waves_3}.
In addition,
shock waves have already been observed
experimentally in the context of rapid potential perturbations related to
the formation of dark solitons \cite{Hau} and vortex lattices
\cite{Shock_JILA}.
In this paper, we discuss both short and long term dynamics in the dimple,
and identify the
distinct growth dynamics induced by the addition of the dimple.
In particular, we highlight the competing effects of temperature, chemical potential and
dimple width on these processes.

We start our analysis by considering the most general scenario of experimental relevance.
The dimple of Fig. 1(a) is added onto a quasi-condensate with about $3,400$ atoms.
This particular dimple has been chosen, as it can be easily generated with existing atom chips.
A detailed discussion of how this can be achieved can be found in Appendix A.


In brief, we find
the subsequent dynamics to be essentially controlled by two parameters:
(i) Firstly, the dynamics depends critically on
the amount of quasi-condensation present in the initial system. This is determined
by the ratio of $T/T_{c}$, where $T_{c}$ is the effective 1D `transition' temperature.
(ii) Secondly, the dynamics is sensitive to the relation of the spatial extent of the atomic cloud
in the original trap compared
 to the effective width of the perturbing dimple potential.

\subsection{Non-equilibrium Dynamics in the Dimple}

For simplicity, we focus initially on the regime $T \ll T_{c}$, for which most
of the atoms are in the quasi-condensate.
Typical density snapshots of the growth dynamics in  this limit
are shown by the black lines in Figs. 2 (a)-(i).
In our analysis, the `transition' temperature
$T_c$ in the presence of interactions is determined numerically from our
modified Popov theory
\cite{Low_D_Theory_2a,Low_D_Theory_2b,Low_D_Theory_2c,Low_D_Theory_2d}.

Once the dimple is turned on,
the central density increases rapidly,
until the
instability towards shock-wave formation is reached
\cite{Shock_Waves_1,Shock_Waves_2,Shock_Waves_3}.
The ensuing shock waves propagate towards the
dimple edge and subsequently reflect back into the dimple centre.
Individual profiles can be simulated by solving the Gross-Pitaevskii equation
for a fully coherent condensate, yielding
density profiles with
large density variations, as shown by the green curves in the insets
of Figs. 2(d)-(h). However,
a detailed understanding of growth dynamics which enables a straightforward comparison to
experiments requires the
addition of random initial phases and subsequent averaging over such
initial configurations. Such averaged profiles, obtained
within our analysis,
reveal the initial formation
of a large central peak, followed by its
break-up into
two smaller peaks, moving in opposite directions
towards the dimple edges, with the central peak
re-emerging after a characteristic time determined
by the dimple potential. The latter peak further splits into
counter-propagating lower peaks, and so forth,
until the gas
equilibrates to the profile shown
in Fig. 2(i).
Such shock-wave dynamics would be largely suppressed
if the dimple was much shallower, or if it were introduced on a much slower timescale.

\begin{figure}[t]
\includegraphics[width=8.cm]{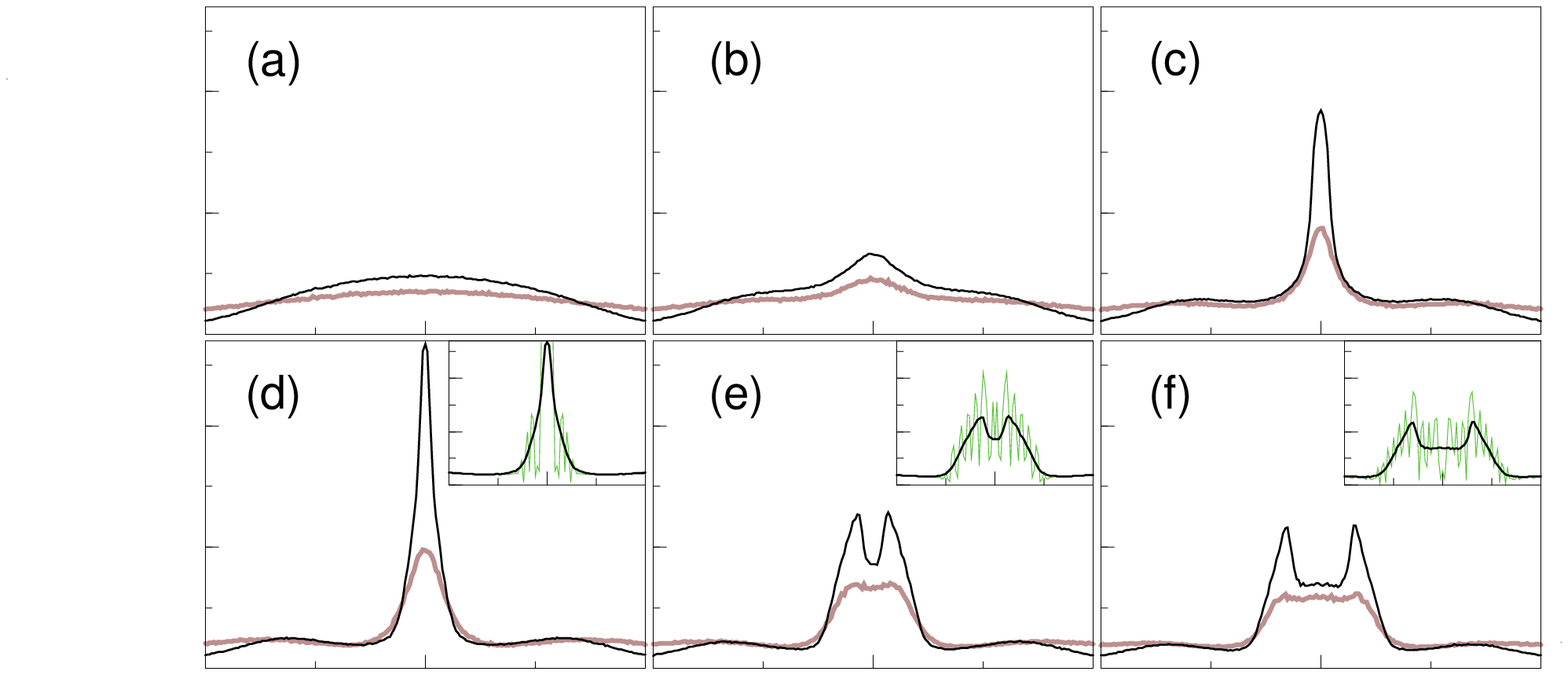}\\
\includegraphics[width=8.cm]{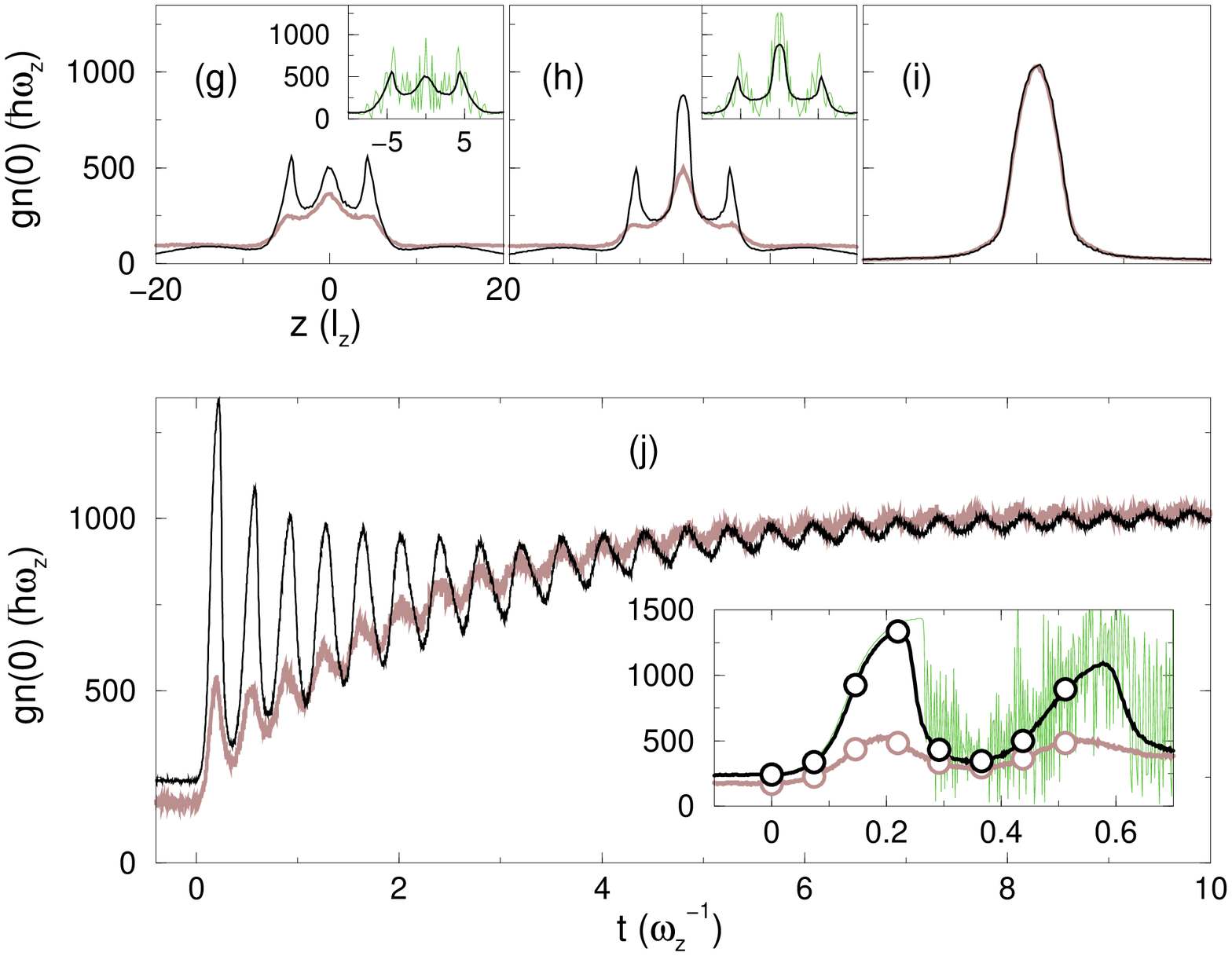}
\caption{
(color online)
Averaged snapshots of total density in the initial harmonic (a) and perturbed ((b)-(i))
trap, taken at equal time intervals, $\Delta t = 0.073 \omega_{z}^{-1}$,
for a gas of 3400 $^{87}$Rb atoms.
Profiles are shown at
$T = 50 {\rm nK} \ll T_{c}$ (black lines) and
$T = 200 {\rm nK} \approx T_{c}$
(brown).
(i) Indistinguishable equilibrium profiles due to large dimple depth.
Insets compare single-run results based on the
Gross-Pitaevskii equation (green)
to averaged stochastic profiles at $T= 50$nK.
(j) Corresponding oscillations in averaged density at the trap centre. Inset highlights
the first two oscillations, showing explicitly the times of the plotted snapshots (a)-(h),
and comparing to results of Gross-Pitaevskii (green).
Dimple parameters as in Fig. 1(a).
}
\end{figure}

The long-term system dynamics is portrayed
in the evolution of the central density
of the gas shown by the black lines in Fig. 2(j).
It features oscillations, on top of a growth curve.
This complicated dynamics is a result of a number of competing processes occuring simultaneously.
Firstly, the observed oscillations are a direct consequence of averaged shock-wave propagation
in the quasi-condensate.
In fact, the initial oscillatory dynamics in the low temperature limit
resembles a suitably averaged single-run of the Gross-Pitaevskii equation,
shown by the green lines in the inset to Fig. 2(j).
The `growth' part of the dynamics arises as a result of spatial compression of the trap,
and features various contributions:
In addition to quasi-condensate  compression,
there is an additional process whereby all
thermal atoms located in the dimple area fall into the perturbed trap by joining
the quasi-condensate.
This is ensured by the
depth of the applied dimple, which largely exceeds all other relevant energies of the system.
Moreover, thermal atoms located in the tails of the initial
atomic cloud (as well as any quasi-condensate atoms located outside of the region of the dimple)
are pulled into the trap centre, continuously interacting with the
propagating quasi-condensate shock wave, and eventually leading to additional
quasi-condensate growth.
In the remaining part of the paper we
discuss these competing processes in
more detail by
identifying suitable experimentally realistic limiting regimes.

\subsection{Oscillation Frequencies}

Firstly, we comment on the observed oscillations.
The addition of the dimple corresponds essentially to an instantaneous local compression of the
central region of the trap.
For an easy visualization which captures the main dynamics discusssed here, let us consider
the simplified case of a harmonic dimple of width comparable to the system size,
such that essentially all of the dynamics is contained within a suitably defined
`dimple region'. In this case, the effect induced by the
addition of the dimple is similar to that of an instantaneous increase
in the harmonic frequency.
This leads to excitation of the lowest compressional mode, which,
for a pure condensate in the 1D mean field regime,
has been predicted to occur at a frequency of
$\sqrt{3} \omega_{D}$ \cite{Oscil_1,Oscil_2},
as already observed experimentally \cite{Oscil_Exp}, and additionally verified numerically in our
simulations.

In the experimentally more realistic gaussian dimples considered here,
we can define an effective harmonic frequency $\omega_{D}=\sqrt{V_{0}}/w$
at the central dimple region.
Although the initial oscillation in the gaussian dimple occurs very close to the
predicted frequency,
the frequency shifts towards lower values,
as soon as the quasi-condensate no longer
feels a purely harmonic confinement, i.e., for system sizes larger than the effective dimple width.
This shift becomes more pronounced
in time as the quasi-condensate grows in the dimple.
In addition, the observed oscillations are damped due to various effects.
Firstly, unlike for a purely coherent system,
the averaging performed over random initial phases in the presence of fluctuations leads to dissipation,
precisely as would be the case in averaged condensate
growth experiments in the presence of a non-negligible thermal cloud.
Furthermore,  the relative motion between the shock waves and the (mostly thermal) atoms entering
the dimple region from the edges is expected to create an additional channel for
dissipation. Finally,
the addition of a gaussian dimple leads to
excitations of multiple frequencies,
which lead to
beating effects and
an accelerated decay of the oscillation amplitude.


\subsection{Effect of Dimple Width}

The above example featured a combination of shock-wave propagation and growth.
These effects can be isolated by changing the form of the perturbation.
Shock waves arise in the quasi-condensate already present prior to the
addition of the dimple.
On the other hand, the observed increase in the central density
is the result of both local
compression of the atoms in the dimple region,
and of direct growth arising from (mainly thermal) atoms
moving into the central region from the trap edge.
As a result, oscillations dominate when the dimple width is comparable to the
quasi-condensate system size,
and for relatively small atom numbers,
such that growth is minimized.
On the other hand, the growth features are dominant in very tight
traps for which the dimple width is much smaller than the system size.

\begin{figure}[t]
\includegraphics[width=8.0cm]{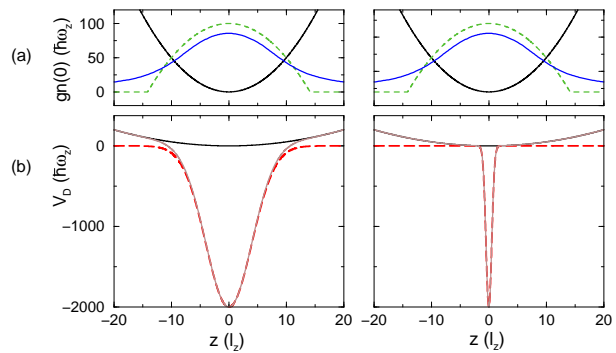}
\caption{
(color online)
(a) Initial density for a gas of $N \approx 960$ atoms at $T=25$nK and $\mu=100 \hbar \omega_{z}$,
such that $\mu/k_{B}T \approx 1 $ (solid blue lines), versus corresponding zero-temperature
Thomas-Fermi profiles (dashed green lines).
The harmonic potential is also shown for easier visualization (solid black).
(b) Initial (solid black) versus final (solid brown) potentials for the two dimples of Fig. 1.
Corresponding dimple potentials $V_{D}$ are shown by the dashed red lines.
}
\end{figure}

These two contrasting regimes for the two opposite limits of wide and tight
dimples, with respect to the effective system size, are shown in Fig. 3
for a reduced number of $N \approx 960$ atoms for computational convenience.
To facilitate an easy comparison between these two limits
we keep the dimple depth fixed.
To give a simple visualization of this distinction, we note that,
for the low temperature case considered here,
the zero-temperature Thomas-Fermi profile (dashed green line in Fig. 3(a)) is
a reasonable first approximation to the system density
prior to the addition of the dimple (solid blue line).
In the chosen harmonic oscillator units,
the zero temperature Thomas-Fermi radius
is defined by $R_{TF}=\sqrt{2\mu}$, where $\mu$ is the chemical potential of the system.

An effective dimple width, $R_{D}$ can also be approximately defined as the point at which
the dimple depth falls to $0.01$ of its maximum value $V_{0}$, implying $R_{D} \approx 3 w$.
Fig. 3 plots both initial densities and potentials (top) and perturbed potentials (bottom)
for the two opposite regimes of $R_{D} \approx R_{TF}$ (left) and $R_{D} \ll R_{TF}$ (right).
Note that the equilibrium densities in both dimple traps are essentially given by the Thomas-Fermi
profiles in these traps. This is true for all atom numbers considered in this paper ($N<3500$), for which
the chosen large dimple depth ensures that all atoms can be accomodated
within the central approximately harmonic dimple region
(see also the density profiles in Fig. 1).

The system dynamics in these two opposite regimes is portrayed by the
evolution of the central density in Fig. 4.
This evolution is plotted both in terms of the original trap timescale $\omega_{z}^{-1}$
in Fig. 4(a), and  in terms of the effective dimple timescale $\omega_{D}^{-1}$
in Fig. 4(b), as these reveal different dynamical features that we wish to comment on.

\begin{figure}[t]
\includegraphics[width=8.cm]{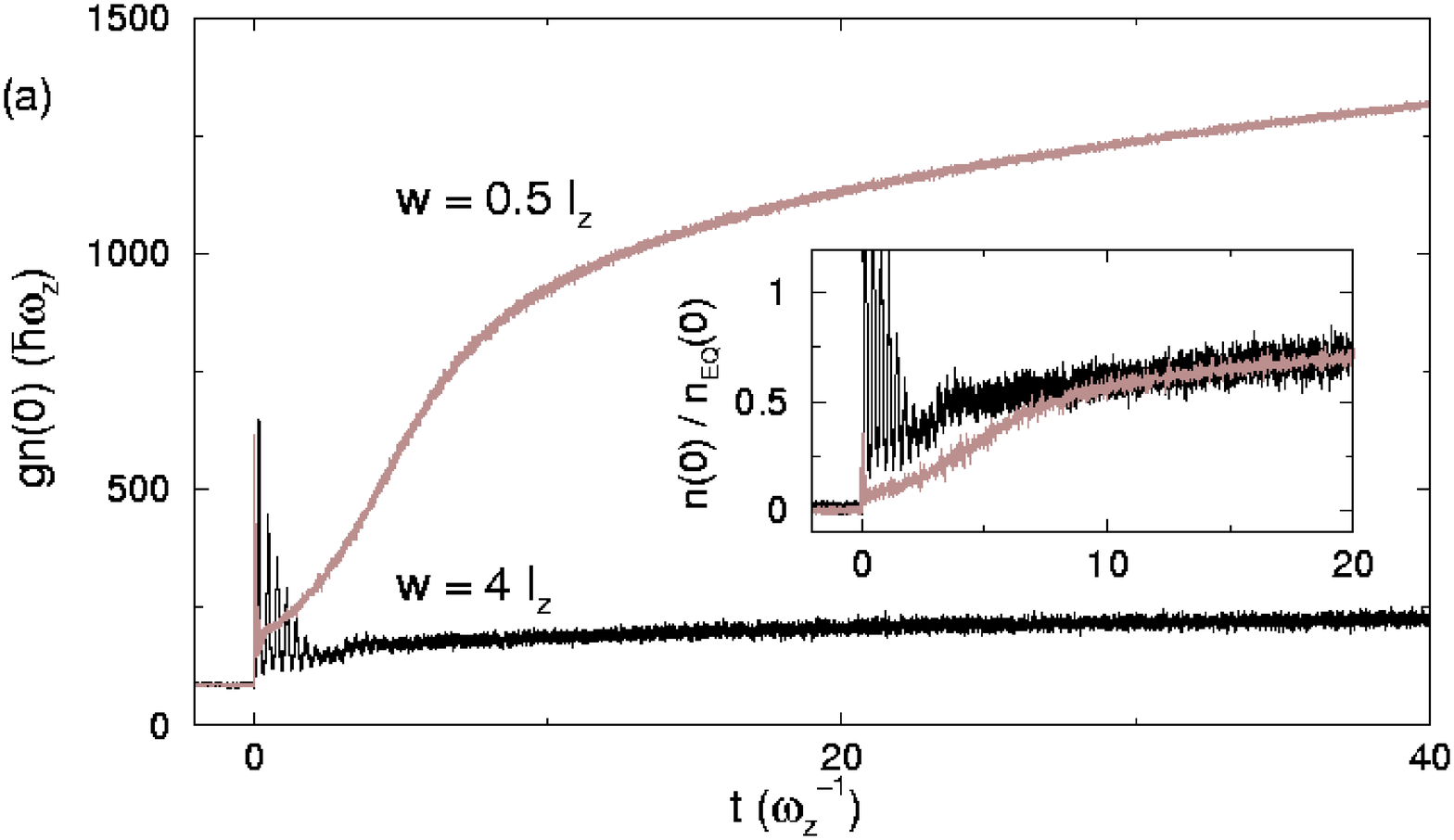}\\
\includegraphics[width=8.cm]{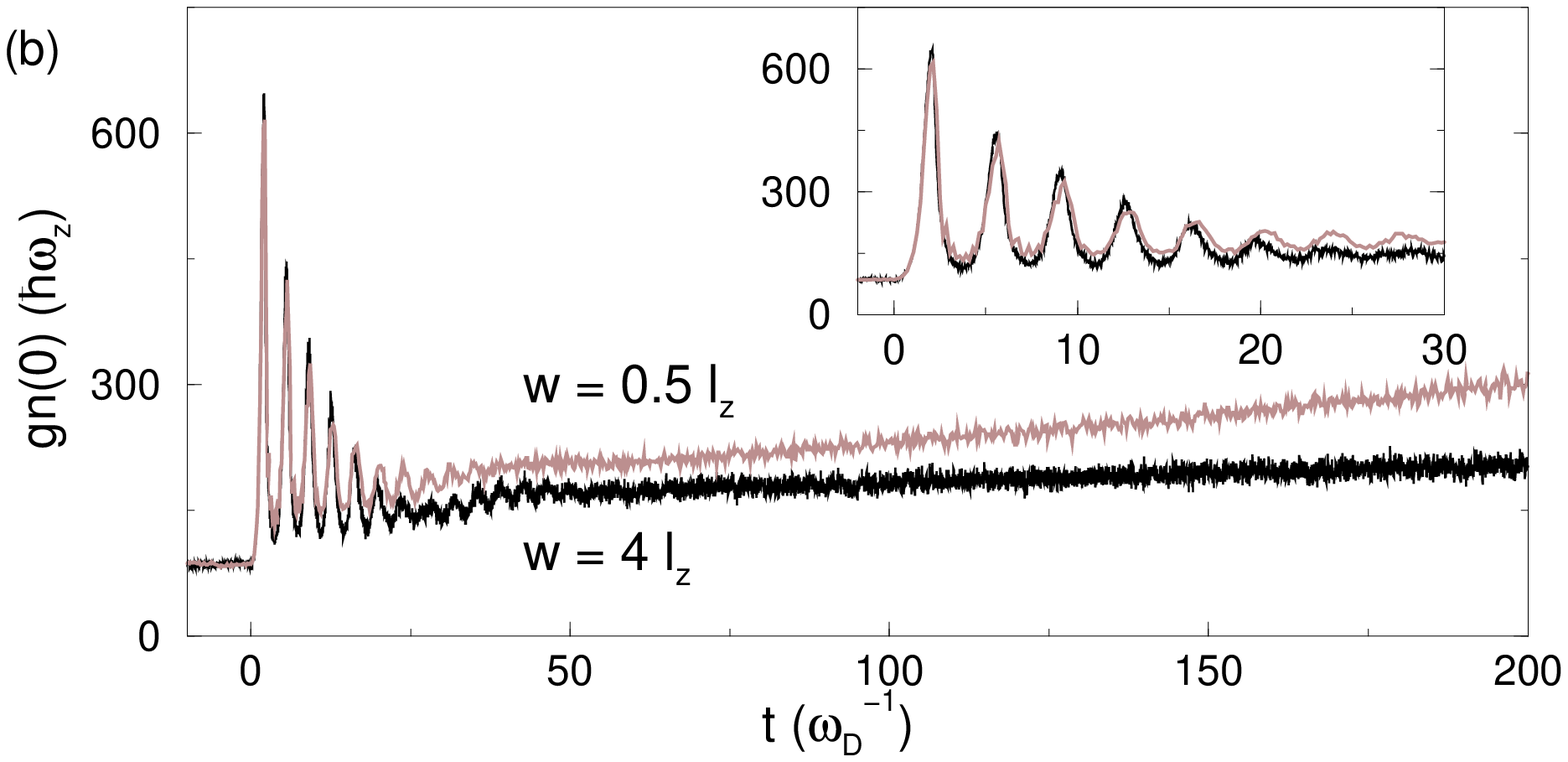}
\caption{
(color online)
(a)-(b) Dynamics of dimple central density for $N \approx 960$ atoms
upon adding a broad dimple with $w=4l_{z}$ (black) or a narrow one with $w=0.5l_{z}$ (brown),
as in Fig. 3.
Time plotted in terms of
(a) original trap timescale $\omega_{z}^{-1}$
and  (b) dimple timescale $\omega_{D}^{-1}$, with
$\omega_{D} = 11.2 \omega_{z}$ (black), or $89.4 \omega_{z}$ (brown).
Here $T=25$nK and $\mu=100\hbar\omega_{z}$, with $\mu/k_{B}T \approx 1 $.
Insets: (a) Same plot with densities approximately scaled to final equilibrium
values $n_{EQ}(0)$, and
(b) Initial regime highlighting the overlapping density oscillations.
}
\end{figure}

The actual relaxation timescale, and long-term dynamics
in these opposing regimes are best compared when both
curves are plotted in terms of
the characteristic timescale $\omega_{z}^{-1}$ of the initial
harmonic trap. Fig. 4(a) thus shows that
the central density in the tight dimple increases much more significantly, due to the
enhanced spatial compression.
A comparison of the growth rate in the different dimple traps is facilitated
in the inset to Fig. 4(a), by plotting
the dynamics of the peak densities, approximately scaled to their respective equilibrium values.
Although equilibration timescales appear to be comparable,
the initial dynamics is
slower in the tight trap,
presumably due to the reduced overlap between the states in the initial and final traps.

The initial dynamics, including the shock-wave-induced oscillations
are more appropriately investigated, when the same results are plotted in terms of the
effective dimple timescale $\omega_{D}^{-1}$, as in Fig. 4(b).
This is because the oscillations are actually
fixed by the frequency $\omega_{D}$ of the perturbed trap (multiplied by the factor
of $\sqrt{3}$), which varies with dimple width $w$.
When plotted in this manner, the initial dynamics  feature
a striking similarity in both amplitude and frequency of the
respective central density oscillations, as shown in the inset.
This is due to the fact that the initial dynamics is set by the quasi-condensate
which is, in both cases, present over the
entire extent of the dimple region.
However, the subsequent growth is noticeably different,
due to the enhanced compression in the tight dimple, leading to
a much enhanced growth in the central density.

\section{Effect of Temperature}

All earlier discussion focused on the regime $T \ll T_{c}$.
We now discuss the effect of changing the initial temperature of the system,
with respect to $T_{c}$.

\subsection{Wide Dimple}

The density snapshots shown in Fig. 2 correspond to fairly `typical' experimental regimes,
for which $R_{TF}(0)$ is approximately equal to a few $R_{D}$,
such that the  addition of the dimple is accompanied by
a combination of oscillations and growth.
However, as the temperature of the initial sample increases
towards $T_{c}$ at fixed atom number, the propagating secondary density peaks
in the averaged profiles become washed out, as shown by the brown lines in Fig. 2(a)-(h).
This is due to the enhanced fluctuations in the initial state prior to the addition of the dimple.
Nonetheless, the
final averaged profile is almost independent of temperature, due to
the large dimple depth, which overshadows
all other
energies of the system.

By comparing these two cases, we note that
the amplitude
of the initial oscillation
is controlled by the ratio of the chemical
potential $\mu$ to the thermal energy $k_{B}T$.
Under conditions of strong condensation, such that
$\mu / k_{B}T > 1$, the initial
oscillation amplitude is large and can
even exceed the equilibrium value.
In the opposite regime, its
amplitude is largely suppressed.
In the limit of
a sufficiently small atom number and a relatively broad dimple, the
addition of the dimple perturbs the central density only mildly.
Thus, the additional constraint of
extremely low temperature ($k_{B}T \ll \mu$), leads
essentially only to the appearance of oscillations without substantial
growth.

\subsection{Tight Dimple}

The tight dimple, whose width is much smaller than the effective system size,
creates a natural temporal separation between the initial shock-wave-induced oscillations and the
subsequent growth dynamics.
These competing effects
exhibit different characteristics, depending on the ratio $T/T_{c}$,
as shown in Fig. 5(a).
This ratio can be controlled either by changing the temperature at constant total atom number,
as discussed for Fig. 2 above,
or by varying the atom number at fixed temperature.
In this section, we choose to discuss the latter, as this enables a further
distinction between spontaneous and stimulated bosonic growth.
Note that, for the small atom numbers
$400<N<1000$ considered here, the condition $R_{TF}(0) \gg R_{D}$ is always satisfied
for the dimple with $w=0.5 l_{z}$.

Regarding the initial oscillations in the central density, we note that,
as the atom number decreases, so does also the ratio of $\mu/k_{B}T$
determining the amplitude of the first oscillation.
This leads to
very strong suppresion when $\mu/k_{B}T \ll 1$, as shown by
the bottom curve in the inset to Fig. 5(a).
After quenching of these initial oscillations, we observe a `secondary growth'
dynamical phase.
For small atom numbers, corresponding to large $T/T_{c}$, we observe a very slow
initial growth in the central density. This will be shown to be consistent with spontaneous
quasi-condensate growth from the thermal cloud.
Increasing the atom number leads to a lower value of
$T/T_{c}$, and to a
significant quasi-condensate fraction
in the initial trap, which yields a faster
initial growth rate in the dimple.

\begin{figure}[t]
\includegraphics[width=8.cm]{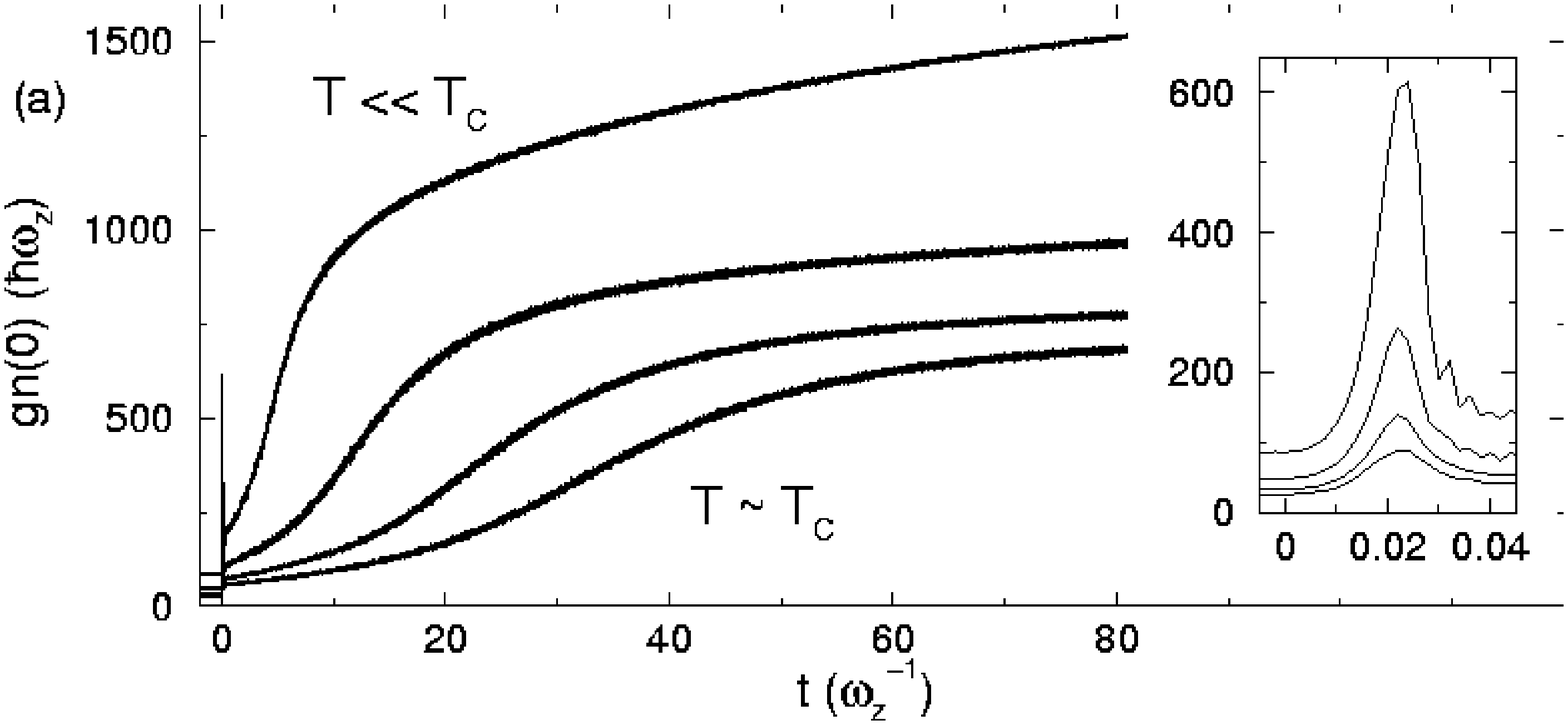}\\
\includegraphics[width=8.cm]{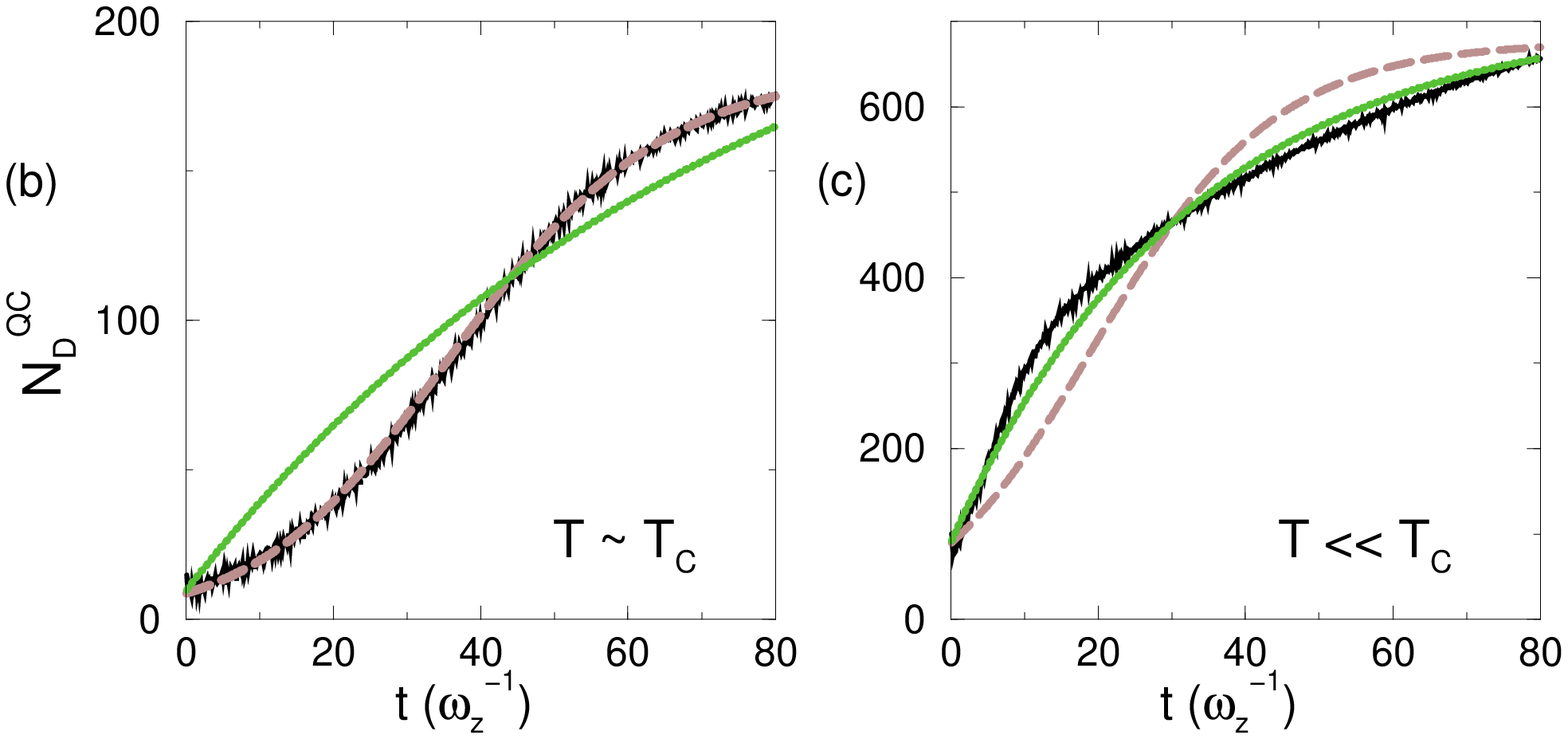}
\caption{
(color online)
(a) Central density dynamics in tight dimple
($w=0.5l_{z}$)
with increasing atom number (from bottom to top) $N\approx$ 420, 490, 620 and 960,
with $R_{D} \ll R_{TF}$ in all cases.
The temperature is fixed to $T=25nK$, and corresponding chemical potentials are
$\mu/\hbar w_{z} = 5, 30, 60, 100$.
Inset: Comparison of initial oscillatory dynamics.
(b)-(c) Quasi-condensate growth in the dimple for
(b) $N \approx 420$ ($T \approx T_{c}$), and
(c) $N\approx960$ ($T \ll T_{c}$).
Each computed curve (black) is fitted with the
growth curve of Eq. (3) (dashed brown), and
the relaxation curve of Eq. (4) (solid green).
The initial number of condensate atoms in the dimple is bigger in (c),
due to the larger chemical potential.
}
\end{figure}


To understand this in more detail, Fig. 5(b)-(c) plots the
integrated quasi-condensate atom number in
the dimple as a function of time for the cases (b) $T \approx
T_{c}$,
and (c) $T \ll T_{c}$. The quasi-condensate atom number is determined
from a combination of local quadratic and quartic correlations of the order parameter
 $\Phi(z,t)$, as will be discussed in more detail elsewhere.
The evolution is in both cases fitted by two distinct growth models,  as in \cite{Ketterle_Growth}.
The first model includes both spontaneous and stimulated bosonic growth, and is fitted by
dashed brown lines. It obeys
\begin{equation}
N_{0}(t)=N_{0}^{(i)}e^{\gamma t}
\left[1+
\left(N_{0}^{(i)}/N_{0}^{(f)}\right)^{\delta}
\left(e^{\delta \gamma t}-1 \right) \right]^{-1/\delta}\;,
\end{equation}
where $N_{0}$ the total quasi-condensate atom number in the dimple,
the superscripts
(i) and (f) denote respectively initial and final values, and $\gamma$ corresponds to
the initial growth rate. The value
$\delta=2/3$ assumes a growth rate linear in the difference of chemical
potential of the thermal cloud and the quasi-condensate, for which $\mu \propto N_{0}^{2/3}$.
The second model, fitted by the solid green lines, describes simple exponential relaxation to equilibrium
via
\begin{equation}
N_{0}(t)=N_{0}^{(f)}\left(1-e^{-\gamma t} \right)\;.
\end{equation}

The limit $T \approx T_{c}$  shown in
Fig. 5(b) is
well described by the growth model of Eq. (3),
indicating spontaneous and subsequent stimulated quasi-condensate growth in the dimple, analogous to
the first condensate growth studies in 3D systems.
Although the exponential relaxation model fails to describe the entire dynamics
in this case, it nonetheless works well if fitted
after a certain `initiation' time
\cite{3D_Growth_Theory_1}.
In
the opposite regime of $T \ll T_{c}$ shown in Fig. 5(c),
the initial dimple dynamics is mainly
governed by re-equilibration of the
perturbed quasi-condensate in the combined trap.
This obeys the exponential relaxation model well, indicating that the dynamics is dominated by stimulated
growth.

The intermediate temperature regime
features interesting, but complicated dynamics, as the effects
of spontaneous and stimulated growth
compete with each other.
The above distinction between spontaneous and stimulated growth becomes less pronounced with
increasing atom number, with the initial growth dynamics being well-modeled by an exponential
relaxation curve even for the tight dimple considered here,
when $N$ is a few $1000$ atoms.

\section{Summary of Distinct Regimes}

The effect of adding a dimple trap to the centre of a weaker harmonic trap containing
a 1D quasi-condensate was considered in the limit when the dimple depth greatly
exceeds all other relevant energies in the system, i.e., $V_{D} \gg \mu, k_{B}T, \hbar \omega_{z}$.
The presented analysis was restricted to the typical experimental regime satisfying
the following length scale separation
\begin{equation}
l_{\perp} < \xi < \left( \lambda_{dB}, l_{D} \right) \ll l_{z} \ll \left( R_{D}, R_{TF}(0) \right)
\nonumber
\end{equation}
where $l_{\perp}=\sqrt{\hbar/m \omega_{\perp}}$ is the transverse harmonic oscillator length,
$\xi = \hbar/\sqrt{4 \mu m}$ is the healing length of the system,
$\lambda_{dB} = \sqrt{2 \pi \hbar^{2} / m k_{B}T}$ is the thermal de Broglie wavelength,
$l_{D}=\sqrt{\hbar/m \omega_{D}}$ is the effective harmonic dimple oscillator length,
$R_{TF}(0)$ is the zero-temperature Thomas-Fermi radius in the original harmonic trap,
and $R_{D}$ is the effective dimple width.
Within this regime, the dynamics in the dimple displays an interesting interplay between
shock-wave propagation in the perturbed quasi-condensate,
and direct quasi-condensate growth.

The following important conclusions were reached about the accessible dynamical regimes in
such systems:\\
\begin{itemize}
\item The initial dynamics for $T < T_{c}$ is dominated by shock-wave propagation,
leading to large oscillations in the central density. The initial amplitude of such oscillations is
controlled by the ratio of $(\mu/k_{B}T)$, yielding large oscillations when this is of
order unity or larger.
\item The long-term system dynamics is dominated by quasi-condensate compression and
growth in the dimple,
with this effect largely pronounced when the system size greatly exceeds the dimple width.
\end{itemize}

\begin{figure}[t]
\includegraphics[width=8.5cm]{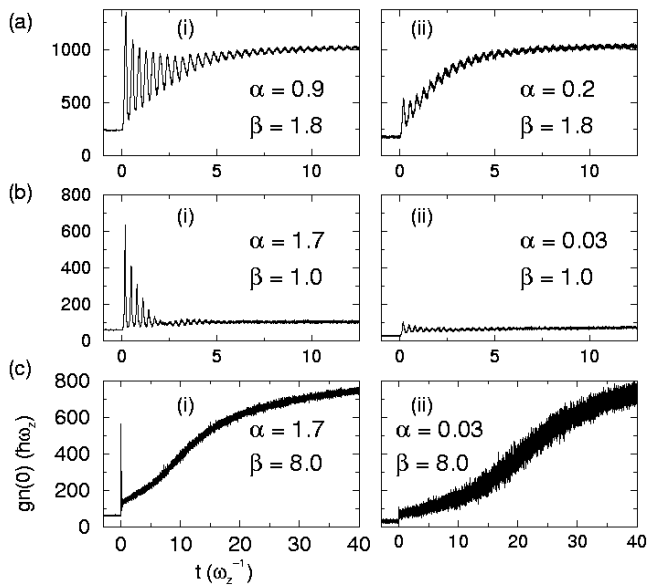}
\caption{Visualization of distinct regimes accessible in ultracold 1D Bose gases,
upon addition of a deep dimple ($V_{0}=-2000 \hbar \omega_{z}$)
at the centre of a weaker harmonic trap (i.e. $z_{D}=0$),
as manifested by the dynamics in the central density
oscillations.
From top to bottom, $N \approx$ (a) 3400, (b)-(c) 540, and $w=$ (a)-(b) $4 l_{z}$,
and (c) $0.5l_{z}$.
Left images (i) correspond to $T \ll T_{c}$, with right images (ii)
showing the opposite regime $T \approx T_{c}$.
Each figure displays the corresponding approximate values of $\alpha = \mu / k_{B}T$, and
$\beta = R_{TF}(0)/R_{D}$, determining the system dynamics.
}
\end{figure}

The different accessible regimes, along with the associated relevant parameters for $(\mu/k_{B}T)$
and $R_{TF}(0)/R_{D}$ are summarized by the evolution of the central density shown in Fig. 6.
In this figure, each horizontal line (a)-(c) corresponds to a particular configuration
of fixed  initial atom number
and dimple width. Left images (i) show the case $T \ll T_{c}$, while
right images (ii) display the high-temperature limit $T \approx T_{c}$.
The images shown portray, from top to bottom, the following cases:
\begin{itemize}
\item[(a)] A relatively large atom number $N=3400$, such that $R_{TF}(0)$ is slightly larger than $R_{D}$.
This leads to noticeable growth, and
additionally features pronounced and `long-lived'
oscillations when $\mu/k_{B}T \approx 1$.
\item[(b)] A sufficiently small atom
number $N=540$ and comparatively wide dimple, $R_{TF}(0) \approx R_{D}$,
for which growth is suppressed and oscillations are the dominant feature in the
low temperature limit.
\item[(c)] A very tight dimple $R_{D} \ll R_{TF}(0)$, featuring `long-term' pronounced
growth, after damping of any initial oscillations occuring.
Note that the spontaneous bosonic growth contribution is pronounced here due to the
small number of atoms.
\end{itemize}
Having identified the importance of the relation of the size of the atomic cloud compared to the dimple width,
 and of the chemical
potential compared to the temperature of the system, we can now `tailor' our experimental conditions to
produce the desired dynamics.

Note that all dimples were considered to have the same depth, which exceeds all other relevant
energy scales in the system. Thus, for the  atom numbers considered here,
the final densities are all described by Thomas-Fermi profiles in the dimple,
and the dimple width is the relevant parameter determining the ensuing 1D dynamics.



The case of an off-centered dimple, with $z_{0} \neq 0$, yields qualitatively
similar behaviour, that will be discussed in more detail elsewhere. When the dimple falls within
the initial quasi-condensate size, shock wave formation plays a key role in the
subsequent dynamics.
However, in the opposite limit of a dimple
located outside of the
quasi-condensate, the
oscillations are largely suppressed, and quasi-condensate relaxation
competes with quasi-condensate growth.
The whole process is further
modified by the asymmetry in the initial density distribution with respect to the
dimple centre, an effect which becomes particularly pronounced for shallow dimples.


\section{Conclusions}

In conclusion, we studied one-dimensional quasi-condensate growth dynamics in a
dimple microtrap created on top of the harmonic confinement
of an atom chip.
Rich novel dynamics were observed in the case
of a deep dimple, whose depth exceeds all other energies
of the system,
displaying an interesting interplay between
shock-wave propagation in the perturbed quasi-condensate,
and direct quasi-condensate growth.
The optimum observation of shock-wave dynamics, without the added significant growth of the
central density, requires low temperatures $T \ll T_{c}$, a comparatively large chemical potential
$\mu > k_{B}T$, and a dimple of width comparable to the system size.
Understanding these processes in more detail is expected to contribute
to fundamental issues in the dynamics of degenerate one-dimensional Bose gases, with 
potential applications
in atom lasers and atom
interferometers.

\acknowledgements
We acknowledge discussions with P. Kr\"{u}ger, S. Wildermuth, and in particular E. Haller.
This work was supported by the Nederlandse Organisatie voor Wetenschaplijk Onderzoek (NWO),
by the European Union, contract
numbers IST-2001-38863 (\emph{ACQP}), MRTN-CT-2003-505032
(\emph{AtomChips}), HPRN-CT-2002-00304 (\emph{FASTNet}), and the
Deutsche Forschungsgemeinschaft, contract number SCHM 1599/1-1.

\appendix
\section{Experimental Creation of 1D Dimple Potentials}

This Appendix discusses various experimental techniques which can be used to
create a dimple potential,
in a quasi-1D atomic gas on an atom chip
\cite{Atom_Chip_Review,fol02,Rei02}, in order to observe the dynamics discussed in this paper.
The basic idea is to modify the 1D trapping
potential with a small but localized attractive potential, which
creates a tight potential minimum along the weak confining axis of
the 1D trap.

To create the situation described in this paper one needs
exceptionally smooth trapping potentials as obtained for example with atom
chips nano-fabricated in gold layers on semiconductor substrates
\cite{fol00,Gro04,Real_1D_2,Prelim_Exp}. On these atom chips, we can routinely
create 1D traps with an aspect ratio of larger than 1000
\cite{wild05a}. The longitudinal confinement in the traps can be
harmonic or box-like on the energy scale of the chemical
potential, depending on the actual setup.

Starting from a 1D potential we can create a dimple on an atom
chip by using
(i) magnetic fields,
(ii) electric fields and
(iii) dipole potentials.
In the following
discussion of creating a dimple we neglect the longitudinal
confinement of the original trap.

\subsection{Creating the dimple with magnetic fields}

\begin{figure}[t]
  \includegraphics[width=8cm]{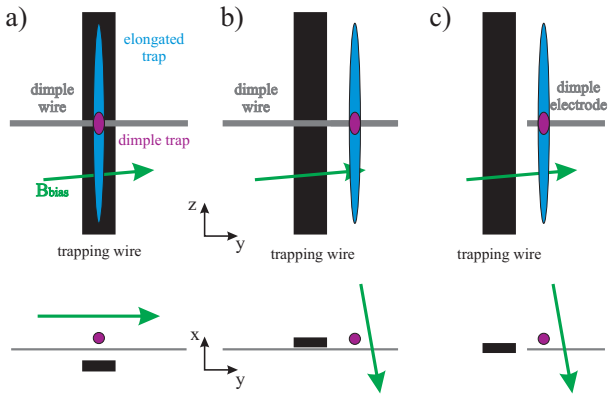}
  \caption{(color online) 
  Wire configurations for creating a dimple in a quasi-1D gas on an atom chip.
  (a) The dimple is created by the crossing wire placed above the trapping wire.
  The magnetic field created by the current flowing in the crossing dimple wire
  subtracts from the minimum of the 1D trap (Ioffe field $B_\textrm{Ioffe}$).
  (b) By rotating the bias field the 1D trap is displaced
  and the dimple is created on the side of the trapping wire.  To
  keep the trap symmetric around the dimple, a wire
  crossing is again required.
  (c) In the rotated geometry charging an electrode creates a strongly confining
  dimple, without the need for wire crossing.
  }\label{Fig:DimpleConfig}
\end{figure}

On an atom chip, the 1D trap is formed in a magnetic minimum
created by superimposing the magnetic field of a current in a
straight wire by a homogeneous bias field
\cite{den99,fol00,Rei01,fol02}. We assume the trapping wire to be
parallel to the \emph{z}-direction (Fig. \ref{Fig:DimpleConfig})
and the homogeneous bias field $B_\textrm{bias}=(B_\textrm{x},
B_\textrm{y}, B_\textrm{Ioffe})$ to be nearly orthogonal, i.e., $\mid
B_\textrm{Ioffe}\mid$ $\ll$ $\mid B_\perp\mid$ with
$B_\perp=(B_\textrm{x},B_\textrm{y},0)$. The component of the
field along the wire direction, the Ioffe field
$B_\textrm{Ioffe}$, defines the minimum of the trapping field.

A dimple can be created by a second wire, located along
the \emph{y}-direction, and thus crossing the trapping wire (Fig.
\ref{Fig:DimpleConfig}) \cite{WireCrossing}.  A small current
$j_\textrm{D}$ in this \emph{dimple}-wire creates a magnetic
field which either \emph{subtracts} from, or \emph{adds} to,
$B_\textrm{Ioffe}$, thus creating a \emph{dimple} or a \emph{barrier} respectively. For
a linear current crossing the trapping wire at $z=0$, with the 1D
trap at height $h$ above it (Fig. \ref{Fig:DimpleConfig} (a)),
the dimple potential is given by
\begin{eqnarray}\label{eq:x}
     V_\textrm{D}(z) =  V_0 \frac{1}{1+(z/h)^2}\;.
\end{eqnarray}
 The depth of the dimple potential
$V_0 \; = \; g_F \mu_B m_F (\mu_0/2 \pi) (j_\textrm{D}/h)$ is determined by
the magnitude of the magnetic field created by the current
$j_\textrm{D}$ at the position of the trap.
Here $\mu_B$ is the Bohr magneton, $g_F$ the Lande factor of the
trapped atomic state $|F,m_F\rangle$ and $\mu_0=4 \pi \; Gmm/A$
is the vacuum permeability.
The longitudinal size (i.e. width) of the dimple is set
by the distance $h$.

The geometry
of the dimple depends on the ratio between  $h$, the height
of the trap above the dimple wire, and
$d_\textrm{trap}$, the
distance of the atoms to the trapping wire:
\begin{eqnarray}\label{eq:md_RatTrapFreq2}
     \frac{\omega_\perp}{\omega_\textrm{D}} &=& \frac{h}{d_\textrm{trap}}
     \frac{\mid B_\perp\mid}{\sqrt{2 B_\textrm{Ioffe} B_\textrm{D}}}
\end{eqnarray}
The parameters $h$ and $d_\textrm{trap}$ can be controlled independently,
by either mounting the dimple wire above the trapping wire (Fig.
\ref{Fig:DimpleConfig}(a)) or by rotating the trap towards the
surface by rotating the bias field (Fig. \ref{Fig:DimpleConfig}(b)),
thus yielding complete control of the dimple trap anisotropy.
When the 1D gas is trapped relatively far from the trapping wire,
the remaining residual disorder potentials
\cite{Real_1D_1,Real_1D_2,Prelim_Exp} are reduced to a point where
they are not visible even for tight transverse confinement of
typical frequency $\omega_\perp \sim 2 \pi (5$ kHz) \cite{Haller04,wild05a}.

In a real experimental setup, the width of the wires and the exact
form of the current distribution in the wires determines the
potential. By measuring the magnetic dimple potential, as described in
\cite{Real_1D_1}, we found that it is actually very well described by a gaussian shape,
as used in the
calculations presented in this paper \cite{Haller04}.

On an atom chip, a 1D trap with $\omega_\perp \sim 2 \pi (5$ kHz)
can be typically created by a $1000$ mA current and a $B_\perp \sim 28G$. 
If the trap is
then rotated towards the chip surface and positioned above an
orthogonal wire at heights $h \sim (5-25) \mu$m, a current
$j_\textrm{D} \sim (10-100) \mu A$ in the dimple wire creates
potentials like the ones described in the main part of this paper.
The larger values of $h$ correspond to wider dimples and require
larger $j_\textrm{D}$.


\subsection{Creating the dimple by adding electric fields}

A second way of implementing a controllable dimple on an atom chip
is by using electric fields (Fig. \ref{Fig:DimpleConfig}(c)).  An
atom with electric polarizability $\alpha$ feels an attractive
potential $V^{{\rm el}}= -\frac{1}{2} \alpha \textbf{E}^2$.  Charging up a
wire positioned orthogonal underneath a trap will create a dimple
potential \cite{Krueger03}.

In a simple model of a 1D trap at height $h$ above the line charge
$Q$, the electric dimple potential
\begin{eqnarray}\label{eq:ElDimple}
     V_\textrm{D}^\textrm{el}(z) = -\frac{1}{2} \alpha \frac{Q}{z^2+h^2}
\end{eqnarray}
has the same form as the magnetic dimple discussed above (Eq.
(\ref{eq:x})). The ratio of the longitudinal to transverse
confinement in the dimple depends again on the distance $d_\textrm{trap}$  between
the trapping wire and the atom trap, and on the height $h$ above
the line charge.  This ratio can thus be adjusted over a broad range of values, by suitably
choosing the position of the trap, as shown in Fig.
\ref{Fig:DimpleConfig}(c).

Such a 1D electric dimple can be easily created on the atom chip
with $\omega_\perp \sim 2 \pi (5$ kHz) described above,
once
%
the trap is rotated towards the
chip surface and positioned
above an orthogonal wire.  
In the electric case, a few Volts of electric potential on
the dimple wire create an adjustable dimple, whose parameters
are similar to the ones described in this paper
for typical trap heights of order $h \sim (5-25) \mu$m
\cite{Haller04}.

In a real setup one must again take into account the detailed
geometry, the exact boundary conditions and the dielectric
properties of the chip substrate. Measuring
the potentials we found that they can be described, to good
accuracy, by a gaussian shape \cite{Haller04}.

Creating the dimple with electric fields has the advantage that
a switchable dimple can be created even without the need for
crossing the wires on the chip, as shown in figure
\ref{Fig:DimpleConfig}(c).  In addition, high
resistivity materials can be used for charged structures, and this will dramatically
reduce the thermal noise induced by Johnson noise currents
\cite{henkel_1,henkel_2,henkel_3,henkel_4,Scheel_1,Scheel_2,Exp_1,Exp_2,Exp_3},
such that distances of 1 $\mu m$ to the charged
dimple wire are feasible.

\subsection{Creating a dimple by superposing a dipole potential}

A third way of creating a dimple on a atom chip is following
\cite{Dimple_Growth_1} and superimposing a dipole potential on the
atom chip trap \cite{Gallego05}.  The dimple is created by a
tightly focused red-detuned laser beam. The size of the dimple is
given by the focus size of the beam.  The form of the dimple
potential is gaussian, $V_\textrm{D} = V_0 \; e^{-(z-z_0)^2/2
\sigma^2}$, and the size of the dimple is set by the numerical
aperture of the imaging lens, and will typically be of order
of 3-10 $\mu m$.

The dipole dimple can be created at arbitrary distance from the
surface of the chip. It is the simplest way to create a tight
dimple in cold atom experiments, and has
been used in the condensate growth experiments at MIT
\cite{Dimple_Growth_1}, and the $^{133}$Cs  experiments at Innsbruck \cite{Dimple_3D_Cs_Exp,2D_Dimple}.
In an atom chip environment, one has to take care
of the stray light from the chip surface and the resulting
interference structure inside the dimple.  Scattered light will
create speckles and form a additional disorder potential.

\subsection{Switching the dimple}

All three examples of dimple traps discussed above can be switched
very fast. In the electric and magnetic cases, only very small
currents or charges are needed, and rapid switching times ($\tau
\ll 1 \mu s$), much faster than the timescale of the transverse
trapping frequency $\omega_\perp$, can easily be obtained. The same
can be said for the dipole dimple, for which the switching depends
only on the light switching time, which can be again $\tau \ll 1
\mu s$.


\begin{thebibliography}{99}

\bibitem{Atom_Chip_Review}
C. Henkel, J. Schmiedmayer, and C. Westbrook, Eur. Phys. J. D {\bf 35} (2005)
Special Issue - Atom Chips: manipulating atoms and molecules
with microfabricated structures.

\bibitem{Chip_Interf_1}
E.A. Hinds, C.J. Vale, and M.G. Boshier, Phys. Rev. Lett. {\bf 86}, 1462 (2001).
\bibitem{Chip_Interf_2}
Y.-J. Wang, D.Z. Anderson, V.M. Bright, E.A. Cornell, Q. Diot, T. Kishimoto,
M. Prentiss, R.A. Saravanan, S.R. Segal, and S. Wu,
Phys. Rev. Lett. {\bf 94}, 090405 (2004).
\bibitem{Chip_Interf_3}
T. Schumm, S. Hofferberth, L.M. Andersson, S. Wildermuth, S. Groth, I. Bar-Joseph,
J. Schmiedmayer, and P. Kr\"{u}ger, Nature Physics {\bf 1}, 57 (2005).
\bibitem{Chip_Interf_4}
Y. Shin, C. Sanner, G.-B. Jo, T.A. Pasquini, M. Saba, W. Ketterle, D.E. Pritchard,
M. Vengalattore, and M. Prentiss, Phys. Rev. A {\bf 72}, 021604 (2005).


\bibitem{Chip_QC_1}
T. Calarco, E.A. Hinds, D. Jaksch, J. Schmiedmayer, J.I. Cirac, and P. Zoller,
Phys. Rev. A {\bf 61}, 022304 (2000).
\bibitem{Chip_QC_2}
P. Treutlein, P. Hommelhoff, T. Steinmetz, T.W. H\"{a}nsch, and J. Reichel,
Phys. Rev. Lett. {\bf 92}, 203005 (2004).
\bibitem{Chip_QC_3}
M.A. Cirone, A. Negretti, T. Calarco, P. Kr\"{u}ger, and J. Schmiedmayer,
Eur. Phys. J. D {\bf 35}, 165 (2005).

\bibitem{Real_1D_1}
S. Wildermuth, S. Hofferberth, I. Lesanovsky, E. Haller, M. Andersson, S. Groth,
I. Bar-Joseph, R. Folman, and J. Schmiedmayer,
Nature {\bf 435}, 440 (2005).
\bibitem{Real_1D_2}
P. Kr\"{u}ger, L.M. Andersson, S. Wildermuth, S. Hofferberth, E. Haller, S. Aigner,
S. Groth, I. Bar-Joseph and J. Schmiedmayer, cond-mat/0504686.





\bibitem{Low_D_Exp_1}
S. Dettmer, D. Hellweg, P. Ryytty, J.J. Arlt, W. Ertmer,
K. Sengstock, D. S. Petrov, G. V. Shlyapnikov,
H. Kreutzmann, L. Santos, and M. Lewenstein,
Phys. Rev. Lett. {\bf 87}, 160406 (2001).
\bibitem{Low_D_Exp_2}
I. Shvarchuck, Ch. Buggle, D.S. Petrov, K. Dieckmann, M. Zielonkowski, M. Kemmann,
T.G. Tiecke, W. von Klitzing, G.V. Shlyapnikov, and J.T.M. Walraven,
Phys. Rev. Lett. {\bf 89}, 270404 (2002).
\bibitem{Low_D_Exp_3}
D. Hellweg, L. Cacciapuoti, M. Kottke, T. Schulte, K. Sengstock, W. Ertmer, and J.J. Arlt,
Phys. Rev. Lett. {\bf 91}, 010406 (2003).
\bibitem{Low_D_Exp_4}
S. Richard, F. Gerbier, J.H. Thywissen, M. Hugbart, P. Bouyer, and A. Aspect,
Phys. Rev. Lett. {\bf 91}, 010405 (2003).


\bibitem{Low_D_Theory_1}
D. S. Petrov, G.V. Shlyapnikov, and J.T.M. Walraven,
Phys. Rev. Lett. {\bf 85}, 3745 (2000).

\bibitem{Low_D_Theory_2a}
J. O. Andersen, U. Al Khawaja,
and H. T. C. Stoof, Phys. Rev. Lett {\bf 88}, 070407 (2002).
\bibitem{Low_D_Theory_2b}
U. Al Khawaja, J.O. Andersen, N.P. Proukakis, and
H.T.C. Stoof, Phys. Rev. A {\bf 66}, 013615 (2002); {\em ibid.} {\bf 66}, 059902(E) (2002).
\bibitem{Low_D_Theory_2c}
U. Al Khawaja,
N.P. Proukakis, J.O. Andersen, M.W.J. Romans, and H.T.C. Stoof, Phys. Rev. A
{\bf 68}, 043603 (2003).
\bibitem{Low_D_Theory_2d}
N.P. Proukakis, Phys. Rev. A {\bf 73}, Vol. 1 (In Press, Jan. 2006) [cond-mat/0505039].

\bibitem{Low_D_Theory_3a}
D.L. Luxat and A. Griffin, Phys. Rev. A {\bf 67}, 043603 (2003).
\bibitem{Low_D_Theory_3b}
C. Mora and Y. Castin, Phys. Rev. A {\bf 67}, 053615 (2003).
\bibitem{Low_D_Theory_3c}
T.K. Ghosh, cond-mat/0402079.
\bibitem{Low_D_Theory_3d}
N.M. Bogoliubov, C. Malyshev, R.K. Bullough, and J. Timonen, Phys. Rev. A {\bf 69}, 023619 (2004).
\bibitem{Low_D_Theory_3e}
D. Kadio, M. Gajda and K. Rzazewski, Phys. Rev. A {\bf 72}, 013607 (2005).


\bibitem{Ketterle_Growth}
H.J. Miesner, D.M. Stamper-Kurn, M.R. Andrews, D.S. Durfee, S. Inouye, and W. Ketterle,
Science {\bf 279}, 1005 (1998).

\bibitem{Dimple_Growth_1}
D.M. Stamper-Kurn,
H.J. Miesner, A.P. Chikkatur, S. Inouye, J. Stenger, and W. Ketterle,
Phys. Rev. Lett. {\bf 81}, 2194 (1998).

\bibitem{3D_Growth}
M K\"{o}hl, M.J. Davis, C.W. Gardiner, T.W. H\"{a}nsch, and T. Esslinger,
Phys. Rev. Lett. {\bf 88}, 080402 (2002).

\bibitem{3D_Growth_Theory_1}
M.J. Bijlsma, E. Zaremba, and H.T.C. Stoof, Phys. Rev. A {\bf 62}, 063609 (2000).
\bibitem{3D_Growth_Theory_2}
C.W. Gardiner, M.D. Lee, R.J. Ballagh, M.J. Davis, and P. Zoller,
Phys. Rev. Lett.
{\bf 81}, 5266 (1998).
\bibitem{3D_Growth_Theory_3}
M.J. Davis, S.A. Morgan, and K. Burnett, Phys. Rev. Lett. {\bf 87}, 160402 (2001).

\bibitem{Compression_1}
W. Ketterle and D.E. Pritchard, Phys. Rev. A {\bf 46}, 4051 (1992).
\bibitem{Compression_2}
P.W.H. Pinkse, A. Mosk, M. Weidem\"{u}ller, M.W. Reynolds, T.W. Hijmans, and J.T.M. Walraven,
Phys. Rev. Lett. {\bf 78}, 990 (1997).

\bibitem{Dimple_3D_Cs_Exp}
T. Weber, J. Herbig, M. Mark, H.-C. N\"{a}gerl, and R. Grimm,
Science {\bf 299}, 232 (2003).

\bibitem{Dimple_3D_Cs_Th}
Z.-Y. Ma, C.J. Foot, and S.L. Cornish, J. Phys. B {\bf 37}, 3187 (2004).

\bibitem{2D_Dimple}
M. Hammes, D. Rychtarik, B. Engeser, H.-C. N\"{a}gerl, and R. Grimm,
Phys. Rev. Lett. {\bf 90}, 173001 (2003).


\bibitem{Prelim_Exp}
P. Kr\"{u}ger, S. Hofferberth, E. Haller, S. Wildermuth, L.M. Anderrson, D. Gallego Garcia,
S. Aigner, S. Groth, I. Bar-Joseph, and J. Schmiedmayer in
L. Marcassa, C. Helmerson and V. Bagnato (eds.),
Atomic Physics 19,
AIP Conference Proceedings {\bf 770}, 144 (2005).

\bibitem{Stoof_Noisy_1}
H.T.C. Stoof, J. Low Temp. Phys. {\bf 114}, 11 (1999).
\bibitem{Stoof_Noisy_2}
H. T. C. Stoof and M.J. Bijlsma, J. Low Temp. Phys.
{\bf 124}, 431 (2001).
\bibitem{Stoof_Noisy_3}
N.P. Proukakis,
Las. Phys. {\bf 13}, 527 (2003).

\bibitem{Olshanii}
M. Olshanii, Phys. Rev. Lett. {\bf 81}, 938 (1998).

\bibitem{Shock_Waves_1}
B. Damski, Phys. Rev. A {\bf 69}, 043610 (2004).
\bibitem{Shock_Waves_2}
A.M. Kamchatnov, A. Gammal, and R.A. Kraenkel,
Phys. Rev. A {\bf 69}, 063605 (2004).
\bibitem{Shock_Waves_3}
V.M. Perez, V.V. Konotop, and V.A. Brazhnyi, Phys. Rev. Lett. {\bf 92}, 220403 (2004).

\bibitem{Hau} Z. Dutton,
M. Budde, C. Slowe, and L.V. Hau,
Science {\bf 293}, 663 (2001).


\bibitem{Shock_JILA}
T.P. Simula, P. Engels, I. Coddington, V. Schweikhard, E.A. Cornell, and R.J. Ballagh,
Phys. Rev. Lett. {\bf 94}, 080404 (2005).

\bibitem{Oscil_1}
S. Stringari, Phys. Rev. A {\bf 58}, 2385 (1998).
\bibitem{Oscil_2}
C. Menotti and S. Stringari, Phys. Rev. A {\bf 66}, 043610 (2002).

\bibitem{Oscil_Exp}
H. Moritz, T. St\"{o}ferle, M. K\"{o}hl, and T. Esslinger,
Phys. Rev. Lett. {\bf 91}, 250402 (2003).




\bibitem{fol02}
    R. Folman, P. Kr\"uger, J. Schmiedmayer, J. Denschlag, and C. Henkel,
Adv. At. Mol. Opt. Phys. \textbf{48}, 263 (2002).

\bibitem{Rei02}
    J. Reichel,  Appl. Phys. B \textbf{74}, 469 (2002).

\bibitem{fol00}
    R. Folman, P. Kr\"{u}ger, D. Cassettari, B. Hessmo, T. Maier, and J. Schmiedmayer,
 Phys. Rev. Lett. \textbf{84}, 4749 (2000).

\bibitem{Gro04}
    S.  Groth, P. Kr\"{u}ger, S. Wildermuth, R. Folman, T. Fernholz, D. Mahalu, I. Bar-Joseph, and J. Schmiedmayer,
Appl. Phys. Lett. \textbf{85}, 2980 (2004).


\bibitem{wild05a}
    S. Wildermuth, PhD. Thesis, Univ. Heidelberg (2005).


\bibitem{den99}
    J. Denschlag, D. Cassettari, and J. Schmiedmayer, Phys. Rev. Lett. \textbf{82}, 2014 (1999)

\bibitem{Rei01}
    J. Reichel, W. H\"ansel, P. Hommelhoff, and T. W. H\"ansch,
    Appl. Phys. B \textbf{72}, 81 (2001).

\bibitem{WireCrossing}
    With the fabrication method described in \cite{Gro04} we have recently
    also fabricated wire crossings at sub-$\mu m$ scale in gold layers and
    semiconductors. (see: www.atomchip.org)

\bibitem{Haller04}
    E. Haller, Diplomarbeit, Univ. Heidelberg (2004).

\bibitem{Krueger03}
    P. Kr\"uger, X. Luo, M.W. Klein, K. Brugger, A. Haase, S. Wildermuth, S. Groth,
I. Bar-Joseph, R. Folman, and J. Schmiedmayer,
Phys. Rev. Lett. \textbf{91}, 233201 (2003).

\bibitem{henkel_1}
     C. Henkel, S. P\"{o}tting, and M. Wilkens, Appl. Phys. B \textbf{69}, 379 (1999).
\bibitem{henkel_2}
     C. Henkel, K. Joulain, R. Carminati, and J.-J. Greffet, Opt. Commun. \textbf{186}, 57 (2000).
\bibitem{henkel_3}
     C. Henkel and S. P\"{o}tting, Appl. Phys. B \textbf{72}, 73 (2001);
\bibitem{henkel_4}
     C. Henkel, P. Kr\"{u}ger, R. Folman, and J. Schmiedmayer, Appl. Phys. B \textbf{76}, 173 (2003).

\bibitem{Scheel_1}
     S. Scheel, P. K. Rekdal, P. L. Knight, and E. A. Hinds, Phys. Rev. A \textbf{72}, 042901 (2005).
\bibitem{Scheel_2}
     P. K. Rekdal, S. Scheel, P. L. Knight, and E. A. Hinds, Phys. Rev. A \textbf{70}, 013811 (2004).

\bibitem{Exp_1}
     M. P. Jones, C. J. Vale, D. Sahagun, B. V. Hall, and E. A. Hinds, Phys. Rev. Lett. \textbf{91}, 080401 (2003).
\bibitem{Exp_2}
     D. M. Harber, J. M. McGuirk, J. M. Obrecht, and E. A. Cornell, J. Low Temp. Phys. \textbf{133}, 229 (2003).
\bibitem{Exp_3}
   Y.-J. Lin, I. Teper, C. Chin, and V. Vuletic Phys. Rev. Lett. \textbf{92}, 050404 (2004).


\bibitem{Gallego05}
    D. Gallego, Diplomarbeit, Univ. Heidelberg (2005).

\end{thebibliography}
\end{document}